\documentclass{svproc}
\usepackage[utf8]{inputenc}
\usepackage{url}

\usepackage[sectionbib]{natbib}
\bibpunct{(}{)}{;}{a}{}{,}%
\usepackage{graphicx}
\usepackage{amsmath,amssymb}
\usepackage{multirow}
\usepackage{longtable}

\usepackage[b5paper]{geometry}
\begin{document}
	\mainmatter              % start of a contribution
	\title{On the influence of multiple stellar populations in globular clusters on 
	their medium-resolution integrated-light spectra}

	\titlerunning{Multiple stellar populations and integrated-light spectra of globular clusters}  % abbreviated title (for running head)
	\author{Margarita Sharina\inst{1} \and Vladislav Shimansky\inst{2}
}
	\authorrunning{Margarita Sharina \and Vladislav Shimansky} % abbreviated author list (for running head)
	\tocauthor{Margarita Sharina, Vladislav Shimansky}
	\institute{Special Astrophysical Observatory, Russian Academy of Sciences,\\
	Nizhnii Arkhyz, 369167, Russia.
		\email{sme@sao.ru},\\ % WWW home page:
	%	\texttt{http://www.sao.ru/hq/sme/index.html}
		\and Kazan Federal University, 18 Kremlyovskaya str., Kazan, 420008, Russia. \\
		\email{Slava.Shimansky@kpfu.ru}
		}
	
	\maketitle

%\begin{document}
%\renewcommand{\ConfSubsection}{Small and medium-size telescopes}
\begin{abstract}
         We take a closer look at our published results of determination of ages, metallicities, helium mass fractions and abundances of chemical elements in Galactic globular clusters in order to find possible signatures of the phenomenon of multiple stellar populations in these data. Our analysis reveals that carbon abundances in the atmospheres of stars in the studied clusters change gradually during their evolution. The changes of the helium mass fraction and C, O, Mg and Na abundance anomalies caused by the effect of multiple stellar populations on the analyzed integrated-light spectra are detected through the comparison of our results with models of chemical evolution and literature data for Galactic field stars.
	\keywords{globular clusters, chemical composition}
\end{abstract}

\section{Introduction}
Light-element abundance anomalies in globular clusters (GCs), i.e. C-N, O-Na and Mg-Al anticorrelations, are signatures of the phenomenon of multiple stellar populations (e.g. Charbonnel, 2016 and references therein). These anomalies do not occur among field stars. The anticorrelations  argue in favor of the existence of several stellar generations in GCs (Gratton et al., 2012). First generation stars were contaminated by the products of p-process nucleosynthesis at high temperatures (Gratton et al. 2012 and references therein). The anticorrelations are
observed in various stellar evolutionary stages from the Main sequence (MS) up to the asymptotic giant branch. Helium mass fractions increase in stars in accordance with the decrease in [C/Fe] and [O/Fe] and with the increase in [N/Fe] (Milone et al., 2018). Maximum helium variation grow with the cluster mass (Milone et al., 2018). Due to the  dredgeup of material processed through the CNO cycle in the globular-cluster red giants, carbon abundances significantly decrease on the way of stars to the red giant branch (RGB) tip (Kraft, 1994 and references therein). 

In this article, we will answer the following question. What information about the variations in the abundances of light elements, including helium, can be extracted from the analysis of medium-resolution integrated-light (IL) spectra of GCs?
\begin{table}
%\scriptsize
	\caption{Literature data for four our sample clusters. Columns 2--4 include data from Harris~(1996): object luminosities in magnitudes, distances in kpc and horizontal-branch ratios, HBR = (B-R)/(B+V+R). Columns 5 and 6 comprise ages in Gyr and metallicities in dex from Sharina et al.~(2020). Columns 7--10 contain average C, O, Mg differences between second and first generation stars and maximum internal helium variations in the clusters from Milone et al.~(2018).}
	\begin{center}
		\begin{tabular}{rccccrcrccccrrccccc}
			\hline \hline
			\multicolumn{1}{c}{\rule{0pt}{12pt}
			Object}& & &\multicolumn{1}{r}{$\rm M_V$} & &\multicolumn{1}{r}
			{$\rm Dist_{\odot}$}& & \multicolumn{1}{r}
			{$\rm HBR$} & & &\multicolumn{1}{r}{$\rm Age$}& & \multicolumn{1}{r}
			{$\rm [Fe/H]$} & & $\rm \Delta [C/Fe]$ & $\rm \Delta [O/Fe]$ & $\rm \Delta [Mg/Fe]$ & & $\rm \delta Y_{max}$\\ [2pt] \hline
			 (1)     & & & (2)  & & (3) & & (4)   & & & (5)  & &  (6) &  & (7)   & (8)   &  (9)  & & (10)  \\ \hline\rule{0pt}{12pt}
			NGC~2808 & & & -9.4 & & 9.6 & & -0.49 & & & 13.6 & & -1.3 &  & -0.75 & -0.60 & -0.25 & & 0.048 \\
			NGC~6205 & & & -8.7 & & 7.7 & & 0.97  & & & 10.0 & & -1.7 &  & -0.15 & -0.35 & -0.10 & & 0.020 \\
			NGC~6723 & & & -7.8 & & 8.7 & & 1.00  & & & 12.6 & & -1.4 &  & -0.35 & -0.39 &  0.00 & & --\\
                        NGC~1851 & & & -8.3 & & 12.1& & -0.36 & & & 12.6 & & -1.6 &  & -0.25 & -0.35 &  0.00 & & 0.007 \\[2pt]
			\hline \hline
		\end{tabular}
	\end{center}
\end{table}

\section{Method}
The details of our approach to population synthesis of IL spectra of GCs were described by Sharina et al. (2020, hereafter: Sh2020). The method was tested in that paper using spectra of 40 Galactic GCs. 
The calculation of synthetic IL spectra of GCs is based on plane parallel hydrostatic models of stellar  atmospheres~(Castelli and Kurucz 2003). The atmospheric parameters were set by isochrones of stellar evolution by Bertelli et al. (2008) with helium content Y between 0.23 and 0.4. The calculated synthetic spectra of individual stars were summed according to the stellar mass function by Chabrier (2005). 
Microturbulence velocities of stars in a cluster were determined by applying a third-degree polynomial function on effective temperature and surface gravity with the coefficients calculated using the data for a sample of 607 stars from the literature (Sh2020). Ages and Y of GCs were determined by comparison of the shapes and intensities of the observed and model absorption lines of the Balmer series of hydrogen and the
intensities of the CaI\,4227, and Ca\,II\,3933.7~\AA, and 3968.5~\AA\ lines. 

To estimate a probable influence of multiple stellar populations on IL spectra and on the derived abundances one has to know 
the contribution of different types of stars to IL spectra of GCs.  
This contribution was estimated by us for NGC~6229 ([Fe/H]$\sim -1.43$~dex, Harris 1996) within the spectral range $3900-5500$~\AA\ 
(Khamidullina et al. 2014, Fig.~14). The contribution of RGB and MS stars looks similar at $\sim\!4800$~\AA. The contribution of other types of stars is lower. At different wavelengths, the contribution of evolutionary stages can vary depending on the age, metallicity and Y of the cluster. Younger stellar populations (hotter MS turnoff point), bluer horizontal branch (HB) stars contribute more the blue part of the spectrum. % The adopted stellar mass function influences synthetic spectra calculation.

\section{Multiple populations in GCs through the analysis of their IL spectra?}
It was demonstrated by Sh2020 that IL carbon abundances are systematically higher by $ \sim\!0.4$ dex than the corresponding mean high-resolution spectroscopic data for red giants in the clusters. This is a consequence of the dredgeup of material processed through
the CNO cycle in red giants (Kraft 1994 and references therein).
\begin{table}
%\scriptsize
	\caption{Helium mass fractions and abundances of C, O, Mg and Na (in dex) with the corresponding 
	typical errors for four GCs selected from the paper by Sh2020 in
	comparison with the model data from Kobayashi et al. (2006) corresponding to [Fe/H]$\sim-1.5$~dex (see text for details).}
		\begin{center}
		\begin{tabular}{rcrccccccccr}	
			\hline  \hline
			\multicolumn{1}{c}{\rule{0pt}{12pt}
			Object}& & &\multicolumn{1}{c}{[C/Fe]} & &\multicolumn{1}{c}
			{[O/Fe]} & &\multicolumn{1}{c}{[Mg/Fe]} & &\multicolumn{1}{c}
			{[Na/Fe]} & &\multicolumn{1}{c}{$\rm Y$}\\[2pt]
			\hline\rule{0pt}{12pt}
			NGC~2808  & & & -0.35    & & 0.10      & & 0.20      & & 0.45     & &  0.30      \\
			NGC~6205  & & & -0.12    & & 0.30      & & 0.20      & & 0.70     & &  0.30      \\
                        NGC~6723  & & & 0.05     & & 0.30      & & 0.40      & & 0.50     & &  0.26       \\    
			NGC~1851  & & & 0.00     & & 0.40      & & 0.51      & & 0.35     & &  0.26       \\
			Errors    & & & $\pm$0.15 & & $\pm$0.3  & & $\pm$0.15 & & $\pm$0.2 & &  $\pm$0.02  \\ \hline
			Model     & & & $\sim$0.0& & $\sim$0.45& & $\sim$0.5 & & $\sim$0.0& & --        \\ [2pt]
			\hline \hline
		\end{tabular}
	\end{center}
\end{table}

Table~1 summarizes observational properties of four Galactic GCs 
from the sample of Sh2020. These clusters are among the brightest in the Galaxy with rather low metallicity and different relative numbers of stars in the red and blue parts of their HBs.
All the objects have a large number of stars in the extreme blue parts of HBs. NGC~1851 and NGC~6723 are so-called 'anomalous' GCs, in which variations of Fe and s-process elements are observed. 
It can be ascertained that the average C, O and Mg differences between second and first generation stars and maximum internal helium variation in the clusters (Milone et al.~2018) 
increase from the bottom to the top of the table.

In Table~2 we list [C/Fe], [O/Fe], [Mg/Fe], [Na/Fe] and Y determined by Sh2020 by analyzing IL spectra.
The typical errors of the IL abundances from Sh2020 are indicated in the fifth row of the table.
The sixth row contains the corresponding approximate abundances in Galactic chemical evolution models by Kobayashi et al. (2006) at [Fe/H]$\sim-1.5$~dex. 
This chemical composition is typical of Galactic field stars with such metallicities (Kobayashi et al. 2006). The lowest Y value is 0.23  (Bertelli et al., 2008).
It can be seen that the deviations from the chemical composition normal for field stars increase from the bottom to the top of the table.
The trend is the same as in Table~1 (columns 7--10).

One can conclude that IL spectroscopic analysis allows us to detect variations in the abundances of light elements in GCs.

\paragraph{Acknowledgments}
The work was supported by the grant RFBR~18-02-00167a. 


%\bibliographystyle{aa}
%\bibliography{template}
\end{document}